\documentclass[reprint,nofootinbib,superscriptaddress,amsmath,amssymb,aps]{revtex4-1}
\usepackage{graphicx}
\usepackage{rotating}
\usepackage{dcolumn}
\usepackage{bm}
\usepackage{color}
\usepackage{mathptmx, textcomp}
\usepackage[latin1]{inputenc}
\usepackage{braket}
\usepackage[colorlinks,linkcolor=magenta,anchorcolor=cyan,citecolor=blue]{hyperref}
\usepackage[multiple]{footmisc}
\newcommand{\fig}[1]{Fig.\ref{#1}}
\def\be{\begin{equation}}
\def\ee{\end{equation}}
\def\ba{\begin{eqnarray}}
\def\ea{\end{eqnarray}}

\def\nn{\nonumber}
\def\lf{\left}
\def\rt{\right}

\newcommand{\eq}[1]{(\ref{#1})}

\def\nn{\nonumber}\def\lf{\left}\def\rt{\right}\def\q{\theta} \def\w{\omega}     \def\p {\pi} \def\a {\alpha}  \def\d {\delta} \def\f {\phi} \def\g {\gamma}    \def\l {\lambda}  \def\x {\xi} \def\c {\chi} \def\b {\beta}  \def\m {\mu} \def\n {\nu} \def\pd {\partial}\def\p {\pi} \def \inf {\infty}  \def \e { \varepsilon}
\def\Q{\Theta}      \def\S {\Sigma}        \def\grad{\nabla}\def\.{\cdot}
\def\math {\mathcal}
\begin{document}

\title{Investigating the gedanken experiment to destroy the event horizon of a regular black hole}
\author{Jie Jiang}
\email{jiejiang@mail.bnu.edu.cn}
\author{Yunjiao Gao}
\email{Corresponding author. 201821140015@mail.bnu.edu.cn}
\affiliation{Department of Physics, Beijing Normal University, Beijing, 100875, China}
\date{\today}

\begin{abstract}
Recently, Li and Bambi proposed a hypothesis that the event horizon of a regular black hole can be destroyed because these objects have no gravitational singularity and therefore they are not protected by the weak cosmic censorship conjecture (WCCC). In this paper, to test their hypothesis, we perform the new version of the gedanken experiments proposed by Sorce and Wald to overcharge a near extremal static electrically regular black hole. After introducing the stability condition of the spacetime and the null energy condition of matter fields, we derive the first-order and second-order perturbation inequalities of the perturbation matter fields based on the Iyer-Wald formalism. As a result, we find that these regular black holes cannot be destroyed under the second-order approximation after these two perturbation inequalities are taken into account, even though they are not protected by the WCCC. Our results indicate that there might be some deeper mechanisms to protect the event horizon of the black holes.

\end{abstract}
\maketitle

\section{Introduction}
A gravitational singularity is usually mathematically ill-defined which leads to the invalidity of predictability of spacetime. Therefore, Penrose formulated the weak cosmic censorship conjecture (WCCC) to ensure the causality of the spacetime. This conjecture states that the gravitational singularity must be hidden inside the event horizon such that it will not be detected by distant observers out of the horizon \cite{RPenrose}. To test the validity of this conjecture, Wald proposed a gedanken experiment to see whether the event horizon of an extremal Kerr-Newman (KN) black hole would be destroyed after throwing a test particle into it \cite{Wald94}. As a result, they found that the extremal KN black holes cannot be destroyed in this way under the first-order approximation of perturbation \cite{Wald94}. However, this experiment depends on two important factors: the initially extremal condition and the first-order approximation. Later, Hubeny extended the discussion to the nearly extremal KN black holes and took the second-order perturbation into account. He showed that the WCCC can be destroyed in this case \cite{Hubeny,1,2,3,4,5}. In the following years, the WCCC received lots of attention and was considered in various gravitational theories \cite{B1, B2, B3, B4, B5,B6,B7,B8,B9,B10,B11,B12,B13,B14,B15,B16,B17}.

However, when considering the second-order approximation, the spacetime cannot be easily treated as a background. We need to consider the full dynamical evolution of the spacetime geometry and the perturbation matter fields. Therefore, Sorce and Wald recently proposed a new version of the gadenken experiment without explicit analysis of the trajectories of the test particles or evolution of the test fields \cite{SW}.  After introducing the null energy condition of the matter fields, they obtained a perturbation inequality for energy, angular momenta, and electric charge of the black hole under the second-order approximation. After this second-order perturbation inequality is taken into account, they showed that these nearly extremal KN black holes cannot be overcharged or overspun at this level.

Recently, this new experiment has also been applied to some other stationary black holes \cite{An:2017phb, Ge:2017vun, Jiang:2019ige, WJ, Jiang:2019vww, Jiang:2019soz, He:2019mqy,Jiang:2020btc} which all showed that the nearly extremal black holes cannot be destroyed under the second-order approximation. We could see that all these black holes have a gravitational singularity inside the event horizon. If there is not a singularity in spacetime, the black holes will not be protected by the WCCC. Therefore, it is natural to ask whether it is possible to destroy a regular black hole.  In 2013, Li and Bambi \cite{Li:2013sea} studied the possibility of destroying the event horizon of regular black holes. They found that the rotating Bardeen and Hayward black holes can be overspun by throwing a test particle under the second-order approximation. Thus they proposed that there are some possibilities to destroy the regular black holes and then distant observers could see inside these black holes. However, their discussion is restricted to the old version of the gedanken experiments. For the story to be truly consistent, it is necessary to extend the investigation into the new version. Moreover, it has been shown in \cite{Rodrigues:2017tfm} that there is some irrationality in a class of the rotating regular black hole solutions, which contain the rotating Bardeen and Hayward black hole solutions. In such cases, we cannot find the Lagrangian density of the nonlinear electromagnetic field. Thus, in this paper, to test their hypothesis, we would like to apply the new version of the gendanken experiments to the static electrically regular black holes and investigate whether the event horizon could also be destroyed under the second-order approximation.

The organization of this paper is as follows: in section \ref{sec2}, we discuss the spacetime geometry of the static electrically regular black holes perturbed by the spherically matter collision. In section \ref{sec3}, based on the Iyer-Wald formalism as well as the null energy condition, we derive the first-order and second-order perturbation inequalities of the perturbation matter fields. In section \ref{sec4}, we perform the new version of the gedanken experiments to destroy the nearly extremal static electrically regular black holes under the second-order approximation of the perturbation. Our conclusions are shown in section \ref{sec5}.
\newpage

\section{electric regular black hole with spherical matter collision}\label{sec2}

In this section, we start by introducing the setup of the new version of the gedanken experiments. The Lagrangian four-form of the static electrically regular black holes is given by
\begin{equation}\label{PL}
	\bm{L}=\frac{\bm{\epsilon}}{16 \pi}\left[R-h(\math{F})\right]+\bm{L}_\text{mt},
\end{equation}
where $\math{F}=F_{ab}F^{ab}$, $\bm{F} = d \bm{A}$ is the strength of the electromagnetic field $\bm{A}$, $R$ is the Ricci scalar related to the metric $g_{ab}$, $\bm{\epsilon}$ is the volume element four-form, and $\bm{L}_\text{mt}$ is the Lagrangian of the perturbation matter fields. The equations of motion can be written as
\ba\begin{aligned}
&R_{ab}-\frac{1}{2}R g_{ab}=8\p \left(T_{ab}^\text{EM}+T_{ab}^\text{mt}\right)\, , \ \ \ \ \ \grad_a G^{ab}=4\p j^b,
\end{aligned}\ea
where
\ba\begin{aligned}
T_{ab}^\text{EM}=\frac{1}{4\p}\left[G_{ac}F_b{}^c-\frac{1}{4}g_{ab}h(\math{F})\right]\,
\end{aligned}\ea
is the stress-energy tensor of the electromagnetic field, and $T_{ab}^\text{mt}$ and $j^a$ are the stress-energy tensor and electric current of the perturbation matter fields, respectively. Here $\bm{G}$ is given by
\ba\begin{aligned}
\bm{G}&=h'(\math{F})\bm{F}\,.
\end{aligned}\ea
The electric charge of the black hole is defined by
\ba\begin{aligned}\label{Qpara}
Q=\frac{1}{4\p}\int_{S_\inf} \star \bm{G}\,.
\end{aligned}\ea
In this paper, we consider the static electrically regular black hole solutions given by \cite{FW}
\begin{equation}
\begin{aligned} \label{metric}
d s^{2} &=-f(r) d t^2+2 dt d r+r^{2}\left(d \theta^{2}+\sin ^{2} \theta d \phi^{2}\right)\,,\\
\bm{A} &=-\frac{q}{\sqrt{2\a}} \left[\left(3-(\m-3)z^\n\right)\left(1+z^\n\right)^{-\frac{\m+\n}{\n}}-3\right]dt\,,
\end{aligned}
\end{equation}
with the blackening factor
\begin{equation}\label{fr}
	f(r)=1-\frac{2M/r}{(1+z^\n)^{\m/\n}}\,,
\end{equation}
where $\n=2,\m, 1$ and $\m\geq 3$. Here we have defined $M=q^3/\a$, $Q=q/\sqrt{2\a}$ and $z=q/r$. The parameters $M$ and $Q$ are the mass and electric charge of the black hole, respectively. Then, the Lagrangian density can be expressed as
\ba\begin{aligned}
h(\math{F})=\frac{2\m}{\a}z^{\m-3}(1+z^\n)^{-\frac{\m+2\n}{\n}}[\m-1-(\n+1)z^\n]\,.
\end{aligned}\ea

In this paper, we would like to consider the situation when all the matter fields satisfy the null energy condition. For the solutions above, it is not difficult to verify that only the case with $\m=3$ (therefore $\n=1,2,3$) satisfies this condition. Therefore we only focus on the case with $\m=3$ from now on. This solution reduces to the Bardeen and Hayward black holes when $\n=2$ and $\n=3$, individually.

Next, we consider a one-parameter family of spherically symmetric matter perturbation to the static electrically regular black holes. Here we assume that the matter collision only occurs in a finite region of the spacetime and the perturbation fields vanish on the bifurcation surface $B$. For simplification, we denote the dynamical fields by $\f(\l)$  in this family, where $\f$ is the collection of $g_{ab}$, $\bm{A}$ and the perturbation matter fields. In this family, the equations of motion is expressed as 
\ba\begin{aligned}
&R_{ab}(\l)-\frac{1}{2}R(\l)g_{ab}(\l)=8\pi\left[T_{ab}^{\mathrm{EM}}(\lambda)+T_{ab}^\text{mt}(\lambda)\right]\,,\\
&\nabla_{a}^{(\lambda)} F^{a b}(\lambda)=4 \pi j^b(\lambda)\\
\end{aligned}\ea
with the condition that $T_{ab}^\text{mt}(0)=j^a(0)=0$ for the background fields. The spacetime in this case can be described by the line element
\ba\begin{aligned}\label{ds2}
ds^2&=-f(r,t,\l)dt^2+2\a(r,t,\l)dr dt\\
&+r^2(d\q^2+\sin^2\q d\f^2)\,,
\end{aligned}\ea
which satisfies $f(r,t,0)=f(r)$ and $\a(r,t,0)=1$. In this case, the background fields are described by Eq. \eq{metric}. With similar consideration to \cite{SW}, we also assume that the spacetime satisfies the stability condition, which states that at sufficiently late times, the spacetime is also described by the electrically regular black hole solution with different parameters which can be labeled by $\l$, which means
\ba
\begin{aligned}\label{S1dsa}
f(r, t, \l)&=f(r,\l)=1-\frac{2M(\l)/r}{[1+z(\l)^\n]^{3/\n}}\,,\\ \a(r,t,\l)&=1\,,\ \ \text{with}\quad z(\l)=q(\l)/r
\end{aligned}\ea
at sufficiently late times.

Due to the stability condition, it is not difficult to see that the stress-energy tensor $T_{ab}^\text{mt}(\l)$ and the electric charge current $j^a(\l)$ vanish at late times. In the above setup, we can introduce a hypersurface $\S=\S_0\cup \S_1$. Here $\S_0$ is a hypersurface with a constant radius $r=r_h$, connecting the bifurcation surface $B$ to a two-dimensional surface $B_1$ at a sufficiently late time. $\S_1$ is a timeslice when $t=t_1$, which connects $B_1$ and an asymptotic sphere $S_\inf$. Then, the dynamical fields on $\S_1$ are described by the expressions in Eq. \eq{S1dsa}.

\section{Perturbation inequalities}\label{sec3}

In this section, following the discussion in \cite{SW}, we would like to derive the first two order perturbation inequalities for the static electrically regular black holes. Since the Lagrangian cannot be  expressed explicitly, below we only consider the off-shell variation of the Lagrangian of the Einstein part, i.e.,
\ba\begin{aligned}\label{action}
\bm{L}=\frac{\bm{\epsilon}}{16\p}R\,.
\end{aligned}\ea
Following the notations in \cite{SW}, we will define
\ba\begin{aligned}
\c=\c(0)\,,\ \ \d\c=\left.\frac{d\c}{d\l}\right|_{\l=0}\,,\ \ \d^2\c=\left.\frac{d^2\c}{d\l^2}\right|_{\l=0}
\end{aligned}\ea
for the quantity $\c(\l)$ in the configuration $\f(\l)$. The off-shell variation of the above Lagrangian gives
\ba\begin{aligned}\label{var1}
\d \bm{L}=\bm{E}_g^{ab} \d g_{ab}+d\bm{\Q}(g,\d g)\,,
\end{aligned}\ea
in which
\ba\begin{aligned}\label{EC}
\bm{E}_g^{ab}&=-\frac{1}{2}\bm{\epsilon}T^{ab}\,,\\
\bm{\Q}_{abc}(g,\d g)&=\frac{1}{16\p}\bm{\epsilon}_{dabc}g^{de}g^{fg}\lf(\grad_g \d g_{ef}-\grad_e\d g_{fg}\rt)\,.
\end{aligned}\ea
Here we have denoted $T_{ab}=T_{ab}^\text{EM}+T_{ab}^\text{mt}$ to the total stress-energy tensor of the electromagnetic field and the perturbation matter fields. The symplectic current three-form is defined by
\ba\begin{aligned}
\bm{\w}(g,\d_1 g,\d_2g)=\d_1\bm{\Q}(g,\d_2g)-\d_2\bm{\Q}(g,\d_1g)\,,
\end{aligned}\ea
and it can be expressed as
\ba\begin{aligned}\label{3w}
\bm{\w}_{abc}&=\frac{1}{16\p}\bm{\epsilon}_{dabc}w^d\,,
\end{aligned}\ea
where
\ba\begin{aligned}
w^a=P^{abcdef}\lf(\d_2g_{bc}\grad_d\d_1 g_{ef}-\d_1 g_{bc}\grad_d\d_2g_{ef}\rt),
\end{aligned}\ea
with
\ba\begin{aligned}
P^{abcdef}&=g^{ae}g^{fb}g^{cd}-\frac{1}{2}g^{ad}g^{be}g^{fc}\\
&-\frac{1}{2}g^{ab}g^{cd}g^{ef}-\frac{1}{2}g^{bc}g^{ae}g^{fd}+\frac{1}{2}g^{bc}g^{ad}g^{ef}\,.
\end{aligned}\ea

Considering a variation corresponding to a diffeomorphism generated by the Killing vector $\x^a=(\pd/\pd t)^a$ of the background spacetime, we can define a Noether current three-form as
\ba\begin{aligned}\label{J1}
\bm{J}_\x=\bm{\Q}(g,\math{L}_\x g)-\x\.\bm{L}\,.
\end{aligned}\ea
According to the discussion in \cite{Wald94}, this current can also be written as
\ba\begin{aligned}\label{J2}
\bm{J}_\x=\bm{C}_\x+d\bm{Q}_\x\,.
\end{aligned}\ea
From Eqs. \eq{var1} and \eq{EC}, we can obtain the constraint $C_\x=\x\.\bm{C}$ with
\ba\begin{aligned}\label{CTJ}
\bm{C}_{dabc}&=\bm{\epsilon}_{eabc}T_d{}^e\,,
\end{aligned}\ea
and the Noether current two-form
\ba\begin{aligned}\label{Q2}
\lf(\bm{Q}_\x\rt)_{ab}&=-\frac{1}{16\p}\bm{\epsilon}_{abcd}\grad^c\x^d\,.\\
\end{aligned}\ea
Based on the above results and the fact that $\math{L}_\x g_{ab}=0$ for the background spacetime, the first-order and second-order variational identities can be derived and they are expressed as
\ba\begin{aligned}
d[\d\bm{Q}_\x-\x\.\bm{\Q}(g,\d g)]&+\x\.\bm{E}_g^{ab}\d g_{ab}+\d \bm{C}_\x=0\,,\\
d[\d^2\bm{Q}_\x-\x\.\d\bm{\Q}(g,\d g)]&=\bm{\w}\lf(g,\d g,\math{L}_\x\d g\rt)\\
&-\d[\x\.\bm{E}_g^{ab}\d g_{ab}]-\d^2 \bm{C}_\x\,.
\end{aligned}\ea
Applying the Stoke's theorem and considering the assumption that the perturbation vanishes on $B$, integration of the first-order variational identity on $\S$ gives
\ba\begin{aligned}\label{var11}
&\int_{S_\inf}\lf[\d\bm{Q}_\x-\x\.\bm{\Q}(g,\d g)\rt]+\int_{\S_1}\x\.\bm{E}_g^{ab}\d g_{ab}\\
&+\int_{\S_1}\d \bm{C}_\x+\int_{\S_0}\d \bm{C}_\x=0\,.
\end{aligned}\ea
For the first term, according to the stability condition, we can calculate it by using the explicit expression \eq{S1dsa} of the dynamical fields at sufficiently late times. So we have
\ba\begin{aligned}\label{dM1}
\int_{S_\inf}\lf[\d\bm{Q}_\x-\x\.\bm{\Q}(g,\d g)\rt]=\d M\,.
\end{aligned}\ea
With a similar calculation, it is easy to verify that
\ba\begin{aligned}\label{Tdg}
T^{ab}(\l)\frac{d g_{ab}(\l)}{d\l}=0
\end{aligned}\ea
on $\S_1$, which implies that the second term of Eq. \eq{var11} vanishes. For the third term, according to Eq. \eq{CTJ}, we can further obtain
\ba\begin{aligned}\label{CL}
\int_{\S_1} \bm{C}_\x(\l)&=-\int_{r_h}^{\inf}\frac{2M(\l)z(\l)^\n[1+z(\l)^\n]^{-\frac{3+\n}{\n}}}{r}dr\\
&=M(\l)\left[\left(1+z_h(\l)^\n\right)^{-3/\n}-1\right]\,,
\end{aligned}\ea
where we have defined $z_h(\l)=q(\l)/r_h$. Then, the third term reduces to
\ba\begin{aligned}
\int_{\S_1} \d\bm{C}_\x(\l)&=\frac{\d M}{(1+z_h^\n)^{3/\n}}-\d M-\frac{3M z_h^\n}{q(1+z_h^\n)^{\frac{\n+3}{\n}}}\d q\,.
\end{aligned}\ea
Combining the above results, we have
\ba\begin{aligned}
\frac{\d M}{(1+z_h^\n)^{3/\n}}-\frac{3M z_h^\n}{q(1+z_h^\n)^{\frac{\n+3}{\n}}}\d q&=-\int_{\S_0}\d \bm{C}_\x\\
&= \d\left[\int_{\S_0}\bm{\tilde{\epsilon}} T_{ab}(dr)^a\x^b\right]\,.
\end{aligned}\ea
In the calculation, we have denoted $\tilde{\bm{\epsilon}}= dt\wedge \hat{\bm{\epsilon}}$ as the volume element on the hypersurface $\S_0$. Next, we are going to investigate the connection between the above variational identity \eq{ineq1} and the null energy condition of the matter fields. According to the expression \eq{ds2} of the spacetime metric, we can choose a null vector field on the hypersurface $\S_0$ as
\ba\begin{aligned}
l^a(\l)=\x^a+\b(\l)r^a\,,
\end{aligned}\ea
in which we have denoted
\ba\begin{aligned}
r^a=\lf(\frac{\pd}{\pd r}\rt)^a\,,\ \ \ \b(\l)=\frac{f(r_h,t,\l)}{2\a(r_h,t,\l)}\,.
\end{aligned}\ea
It is easy to see that $\b=0$ for the background geometry. Then, we can calculate the null energy condition under the first-order approximation of perturbation, that is,
\ba\begin{aligned}
&\int_{\S_0}T_{ab}(\l)l^a(\l)l^b(\l)dt\wedge \hat{\bm{\epsilon}}\simeq\l \d\left[\int_{\S_0}\bm{\tilde{\epsilon}}T_{ab}(dr)^a\x^b\right]\geq 0\,.
\end{aligned}\nn\\\ea
Therefore, the first-order variational identity reduces to
\ba\begin{aligned}\label{ineq1}
\d M-\frac{3M z_h^\n}{q(1+z_h^\n)}\d q\geq0\,.
\end{aligned}\ea
\begin{figure*}
\centering
\includegraphics[width=0.32\textwidth]{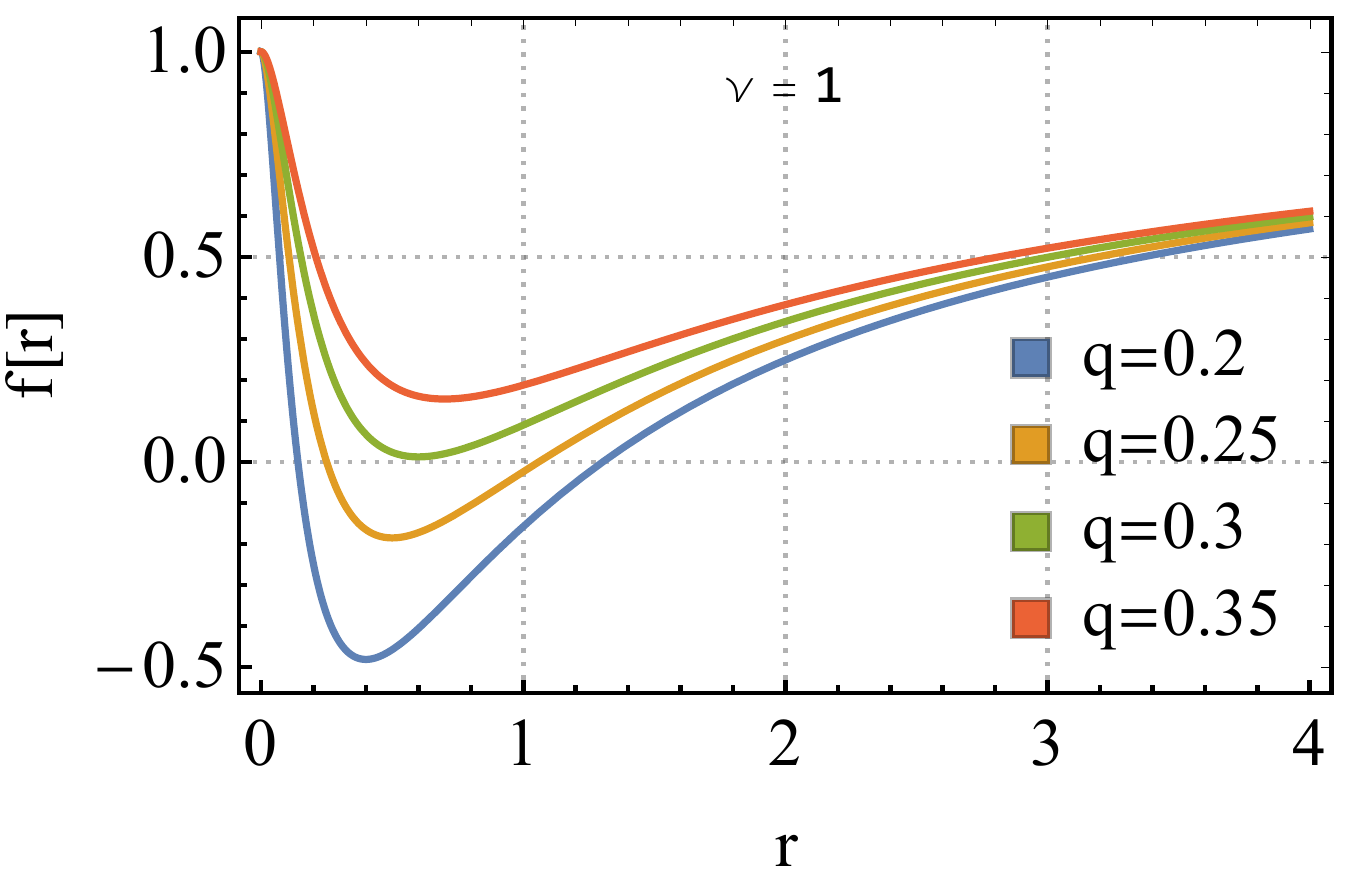}
\includegraphics[width=0.32\textwidth]{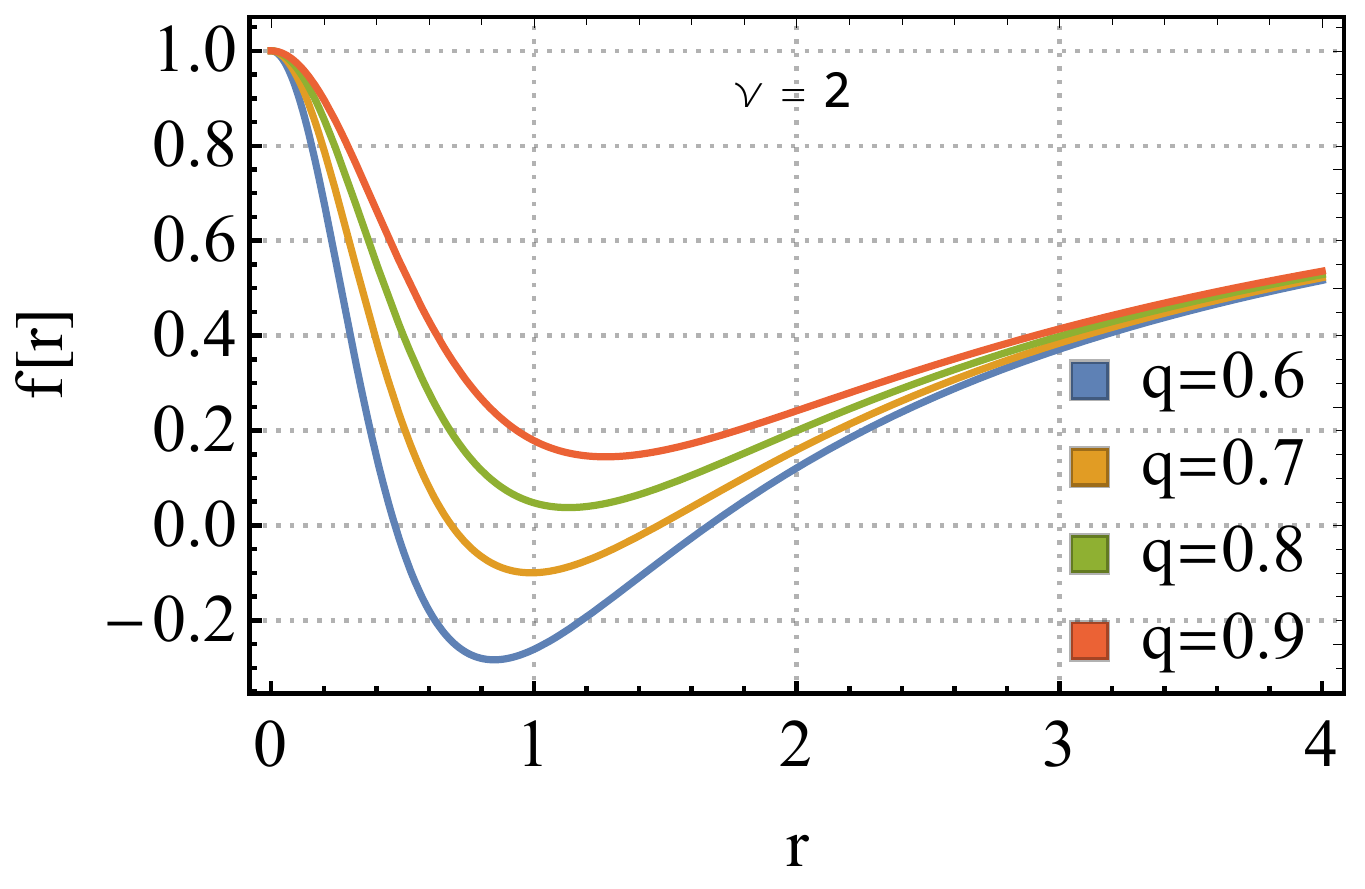}
\includegraphics[width=0.32\textwidth]{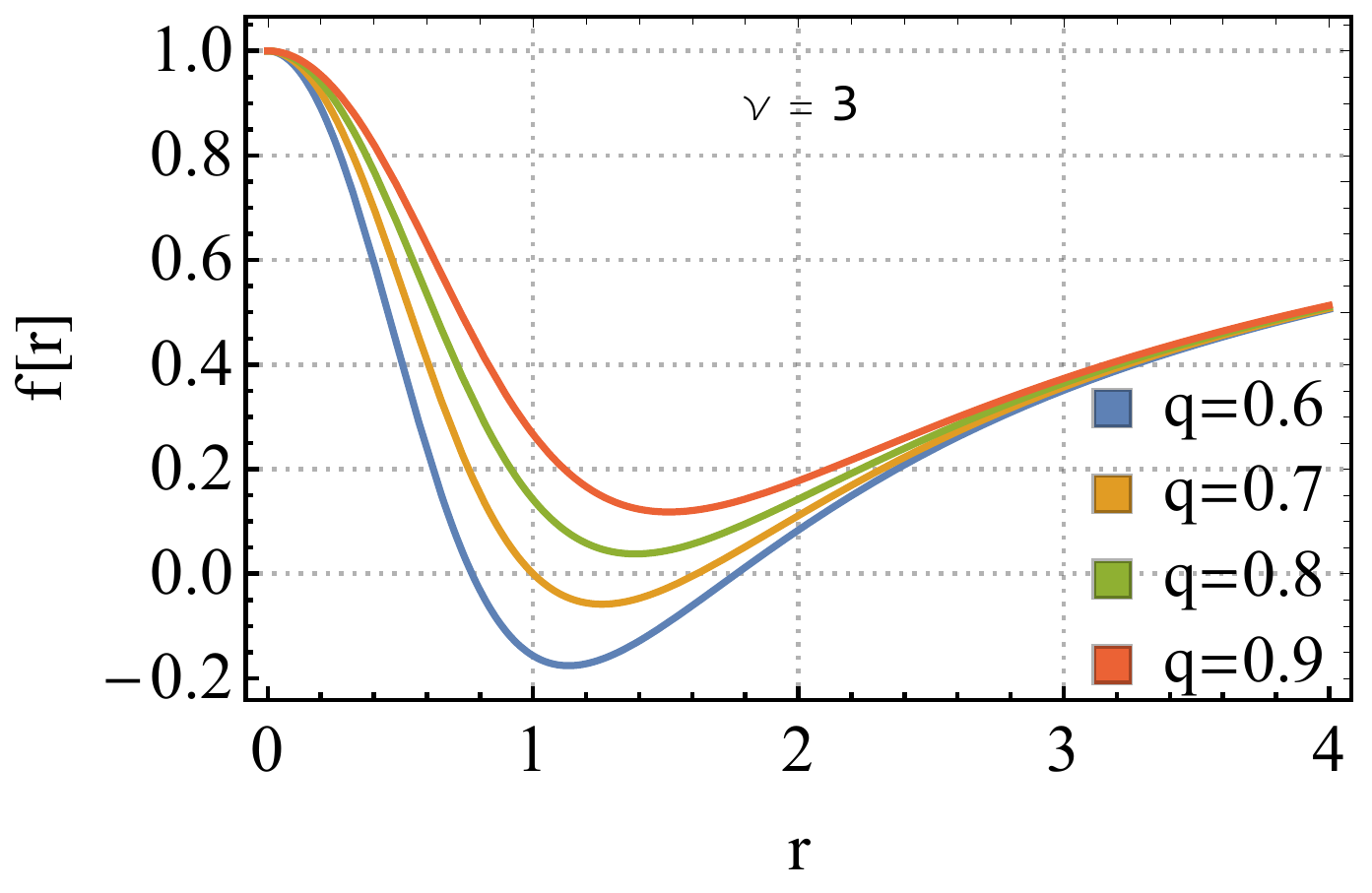}
\caption{Radial dependence of blackening factor $f(r)$ to the electric charge $q$. Here we have set $M=1$.}\label{fig1}
\end{figure*}

As mentioned above, the main purpose of this paper is to investigate whether the above collision process can destroy the event horizon of the static electrically regular black holes. From \fig{fig1}, we can see that the black holes can be destroyed by adding the charge of the spacetime, which means that the optimal choice to violate the black hole configuration is to saturate the first-order perturbation inequality \eq{ineq1}, i.e.,
\ba\begin{aligned}\label{exineq}
\d M-\frac{3M z_h^\n}{q(1+z_h^\n)}\d q=0\,.
\end{aligned}\ea
This also gives $\d\left[\sqrt{-g}T_{ab}(dr)^a\x^b\right]= 0$ on $\S_0$. Straight calculation yields that this condition is equivalent to $\pd_t \d f(r_h,t)=0$, where we  have defined
\ba\begin{aligned}
\d f(r,t)=\left.\frac{\pd f(r,t,\l)}{\pd\l}\right|_{\l=0}\,.
\end{aligned}\ea

For the second-order variational identity in Eq. \eq{var1}, by integrating it on the hypersurface $\S$ and using the Stokes's theorem, we have
\ba\begin{aligned}\label{eq2}
&\int_{S_\inf}\left[\d^2 \bm{Q}_\x-\x\.\d\bm{\Q}(g,\d g)\right]+\int_{\S_1}\d[\x\.\bm{E}_g^{ab}\d g_{ab}]\\
&+\int_{\S_1}\d^2\bm{C}_\x+\int_{\S_0}\d^2\bm{C}_\x-\math{E}_{\S_1}-\math{E}_{\S_0}=0,
\end{aligned}\ea
where we have defined
\ba\begin{aligned}
\math{E}_{\S_i}=\int_{\S_i}\bm{\w}(g,\d g,\math{L}_\x\d g)\,
\end{aligned}\ea
with $i=0,1$. According to Eq. \eq{Tdg} as well as the fact that $\math{L}_\x \f(\l)=0$ on $\S_1$, the second and  fifth terms of the Eq. \eq{eq2} vanish. For the first term, calculating it in a similar way to the first-order identity, we have
\ba\begin{aligned}\label{dM2}
\int_{S_\inf}\left[\d^2 \bm{Q}_\x-\x\.\d\bm{\Q}(g,\d g)\right]=\d^2M\,.
\end{aligned}\ea
For the third term, from Eq. \eq{CL}, we can further obtain
\ba\begin{aligned}
\int_{\S_1}\d^2\bm{C}_\x&=\frac{\d^2 M}{(1+z_h^\n)^{3/\n}}-\d^2 M-\frac{3M z_h^\n}{q(1+z_h^\n)^{\frac{\n+3}{\n}}}\d^2 q\\
&-\frac{3Mz_h^\n(\n-1+2z_h^\n)\d q^2}{q^2(1+z_h^\n)^{\frac{2\n+3}{\n}}}\,,
\end{aligned}\ea
where we have used the optimal first-order perturbation inequality \eq{exineq}. For the last term, based on the explicit expression of the matric \eq{ds2}, we can obtain
\ba\begin{aligned}
\math{E}_{\S_0}&=\frac{r_h}{2}\int_{t_0}^{t_1}dt\left[\pd_t\d\a(r_h,t)\d f(r_h,t)-\d \a(r_h, t)\pd_t \d f(r_h,t)\right]\\
&=\frac{r_h}{2}\d \a(r_h,t_1)\d f(r_h,t_1)=0\,,
\end{aligned}\ea
where $t=t_0$ denotes the coordinate of the bifurcation surface $B$, and
\ba\begin{aligned}
\d \a(r_h,t)=\left.\frac{\pd \a(r_h,t,\l)}{\pd\l}\right|_{\l=0}
\end{aligned}\ea
is the variation of the function $\a(r,t,\l)$. In the above calculation, we have used the optimal condition of the first-order inequality as well as the fact that $\a(r_h,t_1,\l)=1$ on $B_1$.

Summing up the above results, the second-order variational identity becomes
\ba\begin{aligned}\label{secgd}
&\frac{\d^2 M}{(1+z_h^\n)^{3/\n}}-\frac{3M z_h^\n}{q(1+z_h^\n)^{\frac{\n+3}{\n}}}\d^2 q-\frac{3Mz_h^\n(\n-1+2z_h^\n)\d q^2}{q^2(1+z_h^\n)^{\frac{2\n+3}{\n}}}\\
&=-\int_{\S_0}\d^2\bm{C}_\x=\d^2\left[\int_{\S_0}\bm{\tilde{\epsilon}} T_{ab}(dr)^a\x^b\right]\,.
\end{aligned}\ea
Considering the optimal condition of the first-order perturbation inequality, the null energy condition under the second-order approximation gives
\ba\begin{aligned}
\int_{\S_0}T_{ab}l^al^bdt\wedge \hat{\bm{\epsilon}}\simeq\frac{\l^2}{2}\d^2\left[\int_{\S_0}\bm{\tilde{\epsilon}} T_{ab}(dr)^a\x^b\right]\,.\\
\end{aligned}\ea
Then, after taking the null energy condition into account, the second-order variational identity reduces to
\ba\begin{aligned}\label{ineq2}
\d^2 M-\frac{3M z_h^\n}{q(1+z_h^\n)}\d^2 q-\frac{3Mz_h^\n(\n-1+2z_h^\n)\d q^2}{q^2(1+z_h^\n)^{2}}\geq 0\,.
\end{aligned}\ea

\section{Gedanken experiments to destroy the regular black holes}\label{sec4}

In this section, we would like to utilize the new version of the gedanken experiments to overcharge the nearly extremal static electrically regular black holes. Under the assumption of the stability condition, the key point is to check whether the spacetime geometry also describes a black hole at sufficiently late times. Therefore, we define the function
\ba
h(\l)=f(r_m(\l),\l)
\ea
to describe the minimal value of the blackening factor at sufficiently late times. Here $r_m(\l)$ is the minimal radius of the blackening factor, and it can be obtained by
\ba\begin{aligned}
f'(r_m(\l),\l)=0\,,
\end{aligned}\ea
which gives
\ba\begin{aligned}\label{zm}
z_m^\n=1/2\,,\ \ \d r_m=2^{1/\n}\d q\,,
\end{aligned}\ea
with $z_m=q/r_m$. If $h(\l)\leq 0$, the line element at late times describes a black hole geometry. If $h(\l)>0$, there does not exist the event horizon for the late-time geometry and the black hole is destroyed. Under the second-order approximation of perturbation, we have
\ba\begin{aligned}\label{hl0}
h(\l)&\simeq 1-\frac{2M}{q\g}-\frac{2 \l}{q\g}\left(\d M-\frac{M}{q}\d q\right)\\
&-\frac{\l^2}{q\g}\left(\d^2M-\frac{M}{q}\d^2q-\frac{2\d M\d q}{q}+\frac{2M}{q^2}\d q^2\right)\,,
\end{aligned}\ea
where we have set $\g=(27/4)^{1/\n}$. Next, we consider the case where the background spacetime is a nearly extremal black hole, which implies that there is a small parameter $\epsilon$ such that $z_m^\n=(1+\epsilon)z_h^\n$, i.e.,
\ba\begin{aligned}\label{rhrm}
z_h^\n=\frac{1}{2}(1-\epsilon)\,.
\end{aligned}\ea
According to $f(r_h)=0$, one can further obtain
\ba\begin{aligned}\label{eqq1}
M=\frac{1}{2}\g q+\frac{\g q\epsilon^2}{6\n}\,.
\end{aligned}\ea

Following a similar setup to \cite{IW}, we also assume that $\epsilon$ is of the same order to the variational parameter $\l$. According to Eq. \eq{rhrm}, the optimal first-order variational identity can be expressed as
\ba\begin{aligned}\label{eqq2}
\d M-\frac{M}{q}\d q+\frac{2M\d q\epsilon }{3q}=0
\end{aligned}\ea
under the first-order approximation of $\epsilon$. And the second-order perturbation inequality reduces to
\ba\begin{aligned}\label{eqq3}
\d^2 M-\frac{M}{q}\d^2 q-\frac{2\n M \d q^2}{3q^2}\geq 0\,.
\end{aligned}\ea
Utilizing these results, Eq. \eq{hl0} becomes
\ba\begin{aligned}
h(\l)&\leq -\frac{\epsilon^2}{3\n}+\frac{2\d q\epsilon}{3q}-\frac{\n \d q^2}{3q^2}\\
&=-\frac{(q\e-\n\d q)^2}{3q^2\n}\leq 0\,.
\end{aligned}\ea
The above result indicates that the nearly extremal static electrically regular black holes cannot be overcharged by the spherically collision process under the second-order approximation as long as the matter fields satisfy the null energy condition.
\\
\\
\section{Conclusion}\label{sec5}
In this paper, we performed the new version of the gedanken experiments proposed by Sorce and Wald to overcharge a nearly extremal static electrically regular black hole. After introducing the stability conditions of the spacetime as well as the null energy conditions of matter fields, we derived the first-order and second-order perturbation inequalities of the perturbation matter fields based on the Iyer-Wald formalism. As a result, we found that these regular black holes cannot be destroyed under the second-order approximation after these two inequalities are taken into account, even though they are not protected by the WCCC. These results imply that the event horizons may be protected for all of the black holes and there might be some other deeper mechanisms to protect it than the WCCC, such as the second law of the black holes.

\section*{Acknowledgement}
This research was supported by National Natural Science Foundation of China (NSFC) with Grants No. 11775022 and 11873044.
\begin{equation*}
\end{equation*}

\end{document}